\documentstyle[twoside,fleqn,mont,epsfig,epsf]{article}

\newcommand{\as}{$\alpha_s$}
\newcommand{\oas}{$\cal O$($\alpha_s^2$)}

\newcommand{\gev}{\mbox{\,Ge\kern-0.2exV}}
\newcommand{\mev}{\mbox{\,Me\kern-0.2exV}}

\newcommand{\asb}{${\alpha}_0$}
\newcommand{\beq}{\begin{equation}}
\newcommand{\eeq}{\end{equation}}

\newcommand{\AmS}{{\protect\the\textfont2
   A\kern-.1667em\lower.5ex\hbox{M}\kern-.125emS}}

\title{Measurement of $\alpha_s$ and the $\beta$ function with the DELPHI 
detector at LEP}
\author{O. Passon\address{Fachbereich Physik, Bergische Universit{\"a}t 
Wuppertal, Gau\ss{}stra\ss{}e 20, 42097 Wuppertal, Germany}}
\begin{document}

\begin{abstract}
Event shape distributions in $e^+e^-$ annihilation are determined from the 
DELPHI data taken between 183 and 207\gev.
From these the strong coupling $\alpha_s$ is extracted with several 
techniques. Together with the results from other LEP2 
energies and at about $m_Z$ this allows both, a combined measurement of 
$\alpha_s$ and a test of the scale dependence of the strong interaction.
Alternatively the renormalisation  group invariant (RGI) perturbation theory 
is applied to measure the $\beta$ function of strong interaction. The results 
are good agreement with the QCD expectation and allow to exclude the existence
of light gluinos with a mass below 30\gev\ in a model independent way.
\end{abstract}

\maketitle
\section{Introduction}
Neglecting quark masses, $\alpha_s$ is the only free parameter of 
QCD, the theory of strong interaction. Thus its measurement is a task of 
fundamental importance. Additionally, the scale dependence of the strong 
coupling, as governed by the $\beta$ function, is among the important 
predictions of QCD. 

This note presents both, a measurements of $\alpha_s$ at different
LEP2 energies, and a test of the scale dependence of 
the strong interaction. In both cases several techniques are applied.

From the event shapes Thrust, C parameter, heavy jet mass, wide and total jet
broadening, \as\ is extracted with four different methods: The differential 
distributions at each energy are compared to predictions in \oas, pure NLLA 
and matched \oas+NLLA, folded with fragmentation models. Furthermore, \as\ is 
extracted from the mean values using an analytical power correction ansatz. 
These measurements, together with the results from other LEP2 
energies and about $m_Z$, are finally combined into one \as\ 
measurement for each method.

The $\beta$ function is extracted with two different techniques: From the 
\as\ values directly, and from the energy evolution of mean values of event 
shapes. The last method is based on the renormalisation group invariant 
(RGI) perturbation theory \cite{rgi}.

These analyses and the theoretical background are described in more detail 
in \cite{sommer02,sommer02_2} and the references therein.

\section{Data and background}
The analysis is based on data taken with the DELPHI detector
in the years from 1997 to 2000 at centre of mass energies between 183 and 
207\gev. Details on the DELPHI detector and the event selection can be found 
in \cite{sommer02} and the references therein. In total a number of 
$\approx$10.000 hadronic events have been analyses. This is less than 1\% of 
the statistics gathered at $\sqrt{s}=m_Z$. But not only the cross 
section at LEP2 is much smaller than at LEP1. Additionally one has to deal 
with significant background processes. Most important are radiative (ISR) 
events and the production of W pairs (four fermion events). For details we 
refer again to \cite{sommer02}.

\newcommand{\tsppm}{\hspace{\tabcolsep}$\pm$\hspace{\tabcolsep}}
\section{Determination of \boldmath\as\  from event shape distributions 
\label{alphas}}
From the differential distributions of 1-T, C parameter, 
$M^2_{\mathrm{h}}/E^2_{\mathrm{vis}}$,
$B_{\mathrm{max}}$ and $B_{\mathrm{sum}}$, \as\ is determined by fitting 
an \as\ dependent QCD
prediction folded with a hadronisation correction to the data.
As QCD predictions \oas, pure NLLA, and the combined \oas+NLLA in
$\log R$-scheme are employed. 

In \cite{siggi} it has been shown that fixing the
renormalisation scale to $\mu^2=E_{\mathrm cm}^2$ results in a
marginal description of the data.
Therefore, the experimentally optimised scales
from \cite{siggi} are used for the
\oas\ fits. For the NLLA and the combined  NLLA+\oas\
fits, $\mu$ is set equal to $E_{\mathrm cm}$.

\subsection{Theoretical uncertainties}
All perturbative QCD predictions introduce uncertainties due to missing 
higher orders.
The  conventional method for its estimation is to consider the 
effect of a renormalisation scale variation. 
This method however has at least
two  drawbacks: Since the scale error is positively correlated with 
$\alpha_s$, this definition produces a bias towards small $\alpha_s$ values. 
Secondly there are indications, that observables calculated from one 
hemisphere only (like the heavy jet mass or $B_{\mathrm{max}}$)  yield less 
reliable results in  the resummation of leading logarithms \cite{einanpc}. 
This should be reflected in 
their theoretical errors. In contrary  the scale variation yields the 
{\em smallest}  uncertainty for the heavy jet mass and especially 
$B_{\mathrm max}$. For this reason a new definition of the theoretical 
uncertainty for the logR prediction was developed.  
It is based on the additional ambiguity which enters the resumed prediction 
from the phase space restriction. These calculations are forced to vanish at 
the kinematic limit $y_{max}$ using the replacement: 
\begin{eqnarray}
L=\ln \frac{1}{y} \rightarrow L=\ln \left[\frac{1}{X y}-\frac{1}{X y_{max}}+1
\right]
\end{eqnarray}
Usually $X=1$ is chosen, although different values for this $X$ scale
introduce formally only subdominant contributions \cite{xskala}. The 
theoretical error is now defined as half of the difference when $X$ is varied 
between $2/3$ and $3/2$. 
The same definition of the theoretical uncertainty has been adopted for the 
pure NLLA prediction.

For the \oas\ calculation the error from the variation 
around the experimentally optimized scales is still used, as in previous 
publications \cite{siggi,daniel_final_draft}. 

In order to avoid the above mentioned bias,  the corresponding scale 
variations are calculated from theory distributions with a fixed value of 
$\alpha_s$.

\subsection{Combination of the results \label{combi}}
For a combination of the $\alpha_s$ results from different observables at the 
same energy a proper treatment of the correlation is mandatory. An average 
value $\bar{y}$ for correlated measurements $y_i$ is:
\begin{eqnarray*}
\bar{y}=\sum_{i=1}^N w_i y_i
\qquad \mbox{with:}\quad 
w_i = \frac{\sum_j (V^{-1})_{ij}}{\sum_{k,l}(V^{-1})_{kl}}
\end{eqnarray*}
The covariance matrix $V$ has an additive structure for each source of 
uncertainty:
$$
V = V^{\mathrm{stat}}+V^{\mathrm{sys.exp.}}+V^{\mathrm{had}}+
V^{\mathrm{scale}}
$$
The statistical component was estimated with Monte Carlo simulation, while the 
correlation of systematic errors is modeled by the minimum overlap assumption:
\begin{eqnarray}
V_{ij}=\min(\Delta_i^2,\Delta_j^2 )
\end{eqnarray}
Here $\Delta_i$ denotes the corresponding error of the observable $i$.

\section{Determination of \boldmath\as\ from mean values with power 
corrections \label{asfrommeans}}

The analytical  power ansatz is used to determine \as\ from mean event shapes.
This ansatz provides an additive term to the perturbative \oas\ QCD
prediction \cite{dw}.
\begin{equation}
\left< f \right> = 
\frac{1}{\sigma_{\mathrm tot}}\int f\frac{df}{d\sigma}d\sigma =
\left< f_{\mathrm pert} \right> + \left< f_{\mathrm pow} \right> 
\label{eq_f}
\end{equation}
Here an additional non--perturbative parameter \asb\ enters 
for the contributions to the event shape below an infrared matching scale 
$\mu_I$. In
order to measure \as\ from individual high energy data the parameter \asb\
has to be determined by a global fit of \as\ and \asb\ to a large set of
measurements at different energies.
\section{Combination of all DELPHI \boldmath\as\ measurements \label{finalas}}
Assuming the validity of the QCD predicted energy dependence of \as, all 
results can be evolved to a reference energy, e.g. $M_Z$, and combined 
to a single \as$(M_Z)$ measurement. We include also the LEP2 results at 
133, 161 and 172\gev\ from \cite{daniel_final_draft}. For \as\ at and around 
$M_Z$ we have reanalyzed  the distributions from 
\cite{daniel_final_draft} for our five observables and combined the results 
using the same  treatment of correlation as described in section \ref{combi}. 
For \as\ from  mean values the measurements of events with reduced 
centre--of--mass energy  between 44 and 76\gev\ \cite{sommer02_2} have been 
included as well. In combining the \as\ results one faces again the problem 
of correlation. Although the measurements at different energies are obviously
statistically independent, the systematic and theoretical errors are not.
Again this part of the covariance matrix was modeled assuming minimum overlap. 
The only difference to the method described in section \ref{combi} is, that
the error of ISR and four fermion background correlates only those 
measurements which have this source of uncertainty in common. 

\begin{figure}[hbt]
\vspace{9pt}
\mbox{\epsfig{file=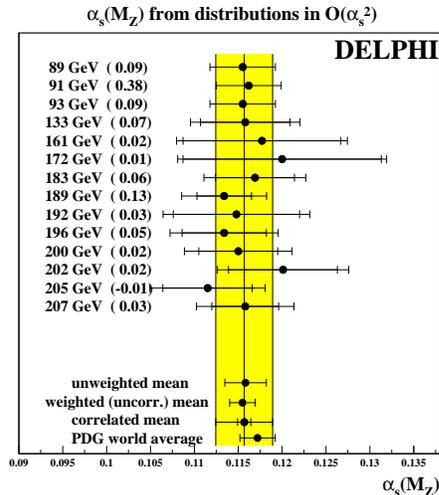,width=7.cm}}\\[-1.7cm]
\caption{Values for $\alpha_s$ at different energies from a fit of the
${\cal{O}}(\alpha_s^2)$ prediction to distributions. Also shown are the results
of different mean value definitions. The numbers in brackets give the 
weights of the individual measurements in the calculation of the correlated 
average. \label{bildchen_as}}
\end{figure}

\begin{figure}[b]
\vspace{9pt}
\mbox{\epsfig{file=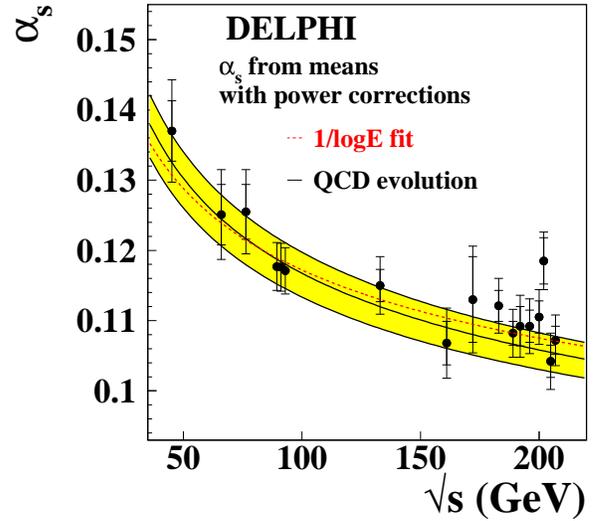,
width=8.5cm}}\\[-1.5cm]
\caption{Scale dependence of $\alpha_s$ from mean values with power 
corrections. The band displays the QCD energy evolution of the mean 
value of this measurements. The dotted line indicates the fit of the scale 
dependence.}
\label{run}
\end{figure}

The result of these combinations are given in Table \ref{final_tab}.
As an example the Figures \ref{bildchen_as} and \ref{run}  
display the combination and energy evolution of the $\alpha_s$ values from the
${\cal{O}}(\alpha_s^2)$ fit to distributions and from mean values with power 
corrections respectively.

As can be seen from Table \ref{final_tab} the total errors are dominated by 
the theoretical uncertainty.  
\begin{table}[t]
\begin{center}
\begin{tabular}{ l c c c }\hline
theory                       & $\alpha_s(M_Z)$ & exp. & theo. \\\hline
${\cal{O}}(\alpha_s^2)$      & 0.1157 & 0.0018 & 0.0027  \\ 
NLLA                         & 0.1093 & 0.0023 & 0.0051  \\ 
${\cal{O}}(\alpha_s^2)$+NLLA & 0.1205 & 0.0021 & 0.0050  \\ 
means        & 0.1184 & 0.0009 & 0.0032  \\ \hline
\end{tabular}
\end{center}
\caption{\label{final_tab}
{  Results of combining all DELPHI \as\ measurements at LEP1 and LEP2.
The quoted errors are the experimental (stat. and sys.exp.) and theoretical 
(scale and hadronisation) ones.}}
\end{table}
\begin{figure}[h]
 \mbox{\epsfig{file=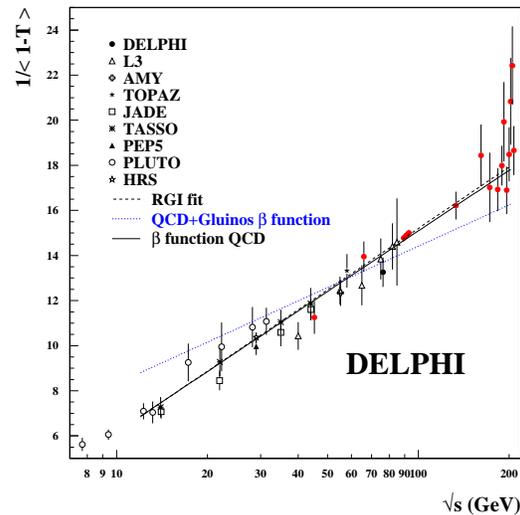,width=7.5cm}
}\\[-1.5cm]
\caption{Scale dependence of 1/$\langle 1-T\rangle$ together with theoretical 
predictions in the RGI framework and the fit result.}
\label{rgi}
\end{figure}
\section{The running of \boldmath\as \label{roas}}
In the previous section we have assumed the validity of QCD in order to 
combine the measurements at different energies. Now we will turn around the 
perspective and use the determination of $\alpha_s$ at different energies 
to {\em test} the  predicted scale dependence of the coupling. The logarithmic
energy slope of the inverse coupling is (in second order) given by:
\begin{eqnarray*}
\label{einspunkt}
\frac{\mathrm{d}\alpha_s^{-1}}{\mathrm{d}\ln Q}=2b_0
+2b_1\alpha_s 
= \left\{\begin{array}{ll}
               1.27&\mbox{(QCD)} \\
               0.89&\mbox{(QCD+gluinos)} \\
              \end{array} \right.
\end{eqnarray*}
Table \ref{betas} gives the slopes when fitting the function
$1/(b\log{\sqrt{s}}+c)$ to the $\alpha_s$ values. Figure \ref{run} displays
the fit result for the $\alpha_s$ values from mean values with power corrections as an example. The results are in good 
agreement with the QCD expectation. The unweighted mean of the different 
measurements yields $1.28\pm 0.12$(RMS). In this fit the correlation between 
the measurements is taken  into account by including the full covariance 
matrix  into the definition of the $\chi^2$ function. The correlation is 
modeled similarly as described in section \ref{combi}. The only differences 
here are that the statistical  errors are uncorrelated, and that the errors 
from ISR and four fermion background correlate only those data which have 
these sources of uncertainty in common. 
\begin{table}[t]
\begin{center}
\begin{tabular}{ll}\hline
theory  & $\frac{d\alpha_s^{-1}}{d\log{E_{\mathrm{CM}}}}$ 
$\pm$ stat $\pm$ sys \\\hline
${\cal{O}}(\alpha_s^2)$             & 1.27 $\pm$ 0.15 $\pm$ 0.33 \\
NLLA                                & 1.40 $\pm$ 0.17 $\pm$ 0.44 \\
${\cal{O}}(\alpha_s^2)$+NLLA  & 1.32 $\pm$ 0.11 $\pm$ 0.27 \\
means (pow.corr.) & 1.11 $\pm$ 0.09 $\pm$ 0.19 \\\hline
\end{tabular}
  \end{center}
\caption{\label{betas}{Results for the slope $b$ when fitting the function 
$1/(b\log{\sqrt{s}}+c)$ to $\alpha_s$ values obtained for the different 
energies.}}
\end{table}
\section{The RGI method}
The $\beta$ functions governs not only the scale dependence of $\alpha_s$, but
directly the energy evolution of e.g. event shapes. 
In the  renormalisation group invariant (RGI) perturbation theory \cite{rgi}
the energy evolution of fully inclusive quantities ${\cal{R}}$, like mean 
values of event shape distributions, can be used to measure the $\beta$ 
function directly:  
$$
Q\frac{dR^{-1}}{dQ}=\frac{\beta_0}{2}+\frac{\beta_1}{4}R+\cdots
= \left\{\begin{array}{ll}
               4.14&\mbox{(QCD)} \\
               2.76&\mbox{(QCD+gluinos)} \\
              \end{array} \right.
$$
Here $R$ is ${\cal{R}}/A$, with $A$ the first order coefficient when expanding
${\cal{R}}$ in powers of $\alpha_s/\pi$. In \cite{sommer02_2} it was shown, 
that this approach yields consistent results for various observables. However,
the highest  accuracy can be achieved for the observable ${\cal{R}}=\langle 
1-T\rangle$ , since here the most low energy data is available.
Our measurement yields:
$$
Q\frac{dR^{-1}}{dQ}=4.21\pm0.18
$$
The fit result is displayed in Figure \ref{rgi}.
This is in good agreement with the QCD expectation and allows to exclude the 
existence of light gluinos in the mass range of $\le$30\gev. 


\end{document}